\documentclass{aastex631}

\begin{document}

\title{An ongoing tidal capture in the Large Magellanic Cloud: the
  low-mass star cluster KMK88-10 captured by the massive globular
  cluster NGC 1835?}\footnote{Based on observations collected at the
Hubble Space Telescope, under proposal GO16361 (PI: Ferraro)}

\author[0000-0002-7717-1022]{Camilla Giusti}
\affiliation{Dipartimento di Fisica \& Astronomia, Universit\`a degli Studi di Bologna, via Gobetti 93/2, I-40129 Bologna, Italy}
\affiliation{INAF - Astrophysics and Space Science Observatory Bologna, Via Gobetti 93/3, 40129, Bologna, Italy }

\author[0000-0002-5038-3914]{Mario Cadelano}
\affiliation{Dipartimento di Fisica \& Astronomia, Universit\`a degli Studi di Bologna, via Gobetti 93/2, I-40129 Bologna, Italy}
\affiliation{INAF - Astrophysics and Space Science Observatory Bologna, Via Gobetti 93/3, 40129, Bologna, Italy }

\author[0000-0002-2165-8528]{Francesco R. Ferraro}
\affiliation{Dipartimento di Fisica \& Astronomia, Universit\`a degli Studi di Bologna, via Gobetti 93/2, I-40129 Bologna, Italy}
\affiliation{INAF - Astrophysics and Space Science Observatory Bologna, Via Gobetti 93/3, 40129, Bologna, Italy }

\author[0000-0001-5613-4938]{Barbara Lanzoni}
\affiliation{Dipartimento di Fisica \& Astronomia, Universit\`a degli Studi di Bologna, via Gobetti 93/2, I-40129 Bologna, Italy}
\affiliation{INAF - Astrophysics and Space Science Observatory Bologna, Via Gobetti 93/3, 40129, Bologna, Italy }

\author[0000-0001-9545-5291]{Silvia Leanza}
\affiliation{Dipartimento di Fisica \& Astronomia, Universit\`a degli Studi di Bologna, via Gobetti 93/2, I-40129 Bologna, Italy}
\affiliation{INAF - Astrophysics and Space Science Observatory Bologna, Via Gobetti 93/3, 40129, Bologna, Italy }

\author[0000-0002-7104-2107]{Cristina Pallanca}
\affiliation{Dipartimento di Fisica \& Astronomia, Universit\`a degli Studi di Bologna, via Gobetti 93/2, I-40129 Bologna, Italy}
\affiliation{INAF - Astrophysics and Space Science Observatory Bologna, Via Gobetti 93/3, 40129, Bologna, Italy }

\author[0000-0003-2742-6872]{Enrico Vesperini}
\affil{Department of Astronomy, Indiana University, Bloomington, IN, 47401, USA}

\author[0000-0003-4237-4601]{Emanuele Dalessandro}
\affil{INAF - Astrophysics and Space Science Observatory Bologna, Via Gobetti 93/3, 40129, Bologna, Italy }
  
\author[0000-0001-9158-8580]{Alessio Mucciarelli}
\affiliation{Dipartimento di Fisica \& Astronomia, Universit\`a degli Studi di Bologna, via Gobetti 93/2, I-40129 Bologna, Italy}
\affiliation{INAF - Astrophysics and Space Science Observatory Bologna, Via Gobetti 93/3, 40129, Bologna, Italy }

\begin{abstract}
In the context of a project aimed at characterizing the dynamical evolution of old globular clusters in the Large Magellanic Cloud, we have secured deep HST/WFC3 images of the massive cluster NGC 1835. In the field of view of the acquired images, at a projected angular separation of approximately 2 arcmin from the cluster, we detected the small stellar system KMK88-10. The observations provided the deepest color-magnitude diagram ever obtained for this cluster, revealing that it hosts a young stellar population with an age of 600-1000 Myr. The cluster surface brightness profile is nicely reproduced by a King model with a core radius $r_c=4\arcsec (0.97$ pc), an half-mass radius $r_{hm}=12\arcsec (2.9$ pc), and a concentration parameter $c\sim1.3$ corresponding to a truncation radius $r_t \sim81\arcsec (19.5$ pc). We also derived its integrated absolute magnitude ($M_V=-0.71$) and total mass ($M\sim80-160 M_\odot$). The most intriguing feature emerging from this analysis is that KMK88-10 presents a structure elongated in the direction of NGC 1835, with an intracluster over-density that suggests the presence of a tidal bridge between the two systems. If confirmed, this would be the first evidence of a tidal capture of a small star cluster by a massive globular.
\end{abstract}

\keywords{Magellanic Clouds; globular clusters: individual (NGC1835); star clusters: individual (KMK88-10); techniques: photometric}

\section{Introduction}
\label{sec:intro}
The Large Magellanic Cloud (LMC) is the closest (irregular) galaxy to the Milky Way (MW). Due to its proximity ($\sim 50$ kpc, \citealt{harris+2009, pietrzynski+2019}), it is possible to resolve individual stars, also within its stellar clusters.  These are both rich star clusters, with properties similar to those of the MW globular clusters, and lower-mass systems resembling the Galactic open clusters. At odds of their Galactic counterparts, globular clusters in the LMC exhibit a wide range of ages (from a few million, to several billion years), thus providing an unique laboratory to study the major changes in the observational properties of stellar populations as a function of the age (see, e.g., \citealt{ferraro+1995, ferraro+2004, mucciarelli+2006}). Moreover, they provide deep insights into the LMC star formation history \citep{baumgardt+2013} and chemical enrichment \citep[e.g.][]{pietrzynski+2000, glatt+2010}.  While these systems have been targeted for several studies through the years \citep[e.g.][]{olsen+1998, brocato+1996, mackey+2003, ferraro+2006, mucciarelli+2008, mucciarelli+2010, ferraro+2019, lanzoni+2019, cadelano+2022a}, an extensive and comprehensive study of LMC low-mass clusters is still lacking. This is because they are composed of a limited number of stars embedded in the LMC field population, which severely hinders their detection and makes them easily misidentifiable with asterisms \citep{bica+2008, choudhury+2015, nayak+2016}. Nevertheless, such systems are crucial for understanding the formation and evolution of the LMC \citep{piatti+2012, palma+2013, piatti+2014, choudhury+2015, palma+2016}.
 
Interestingly, ``star cluster pairs'' with small projected relative distances are frequently observed in the LMC \citep{priyatikanto+2019a, priyatikanto+2019b}. \citet{bhatia+1988, bhatia+1991, subramaniam+1995, dieball+2002} used statistical arguments to prove that the large number of observed cluster pairs cannot be fully explained by random projection effects. This indicates that at least some of them are likely gravitationally bound, and observational confirmations of this hypothesis have been found in a few cases \citep[e.g][]{mucciarelli+2012, desilva+2015, dalessandro+2018}. Interacting binary clusters could be the result of the primordial fragmentation of a common molecular cloud, or the outcome of a close encounter between two clusters during their motion in the galaxy, and a subsequent tidal capture \citep[see, e.g.][]{dieball+2002, mucciarelli+2012, dalessandro+2018}. While in
the former case binary clusters are expected to share some common property (such as the age and chemical composition), in the latter case the two members can have very different ages, masses and chemical abundances.

Cluster pairs generated by encounter processes can bring a wealth of information about the evolutionary history of the host galaxy, and also explain the properties of individual systems with peculiar characteristics, such as unusual rotation patterns \citep{lee+1999, baumgardt+2003, bruns+2011}, or spreads in chemical abundances \citep[e.g.][]{amaroseoane+2013, gavagnin+2016, hong+2017}.
%For this reason binary clusters composed of two approximately equal mass systems have been extensively studied in the past, mainly through \textit{N}-body simulations \citep{portegies+2007, delafuente+2010}, but also through observational analysis \citep{mucciarelli+2012, desilva+2015, dalessandro+2018}. 
For this reason they have been extensively studied in the past, mainly through \textit{N}-body simulations \citep{rao+1987, deoliveira+1998, deoliveira+2000, portegies+2007, delafuente+2010, priyatikanto+2016, darma+2021}. The fate of these systems seems to be twofold: those having a small initial separation will likely experience a merger in a relative short timescale, while the ones with larger initial separation will probably be moved apart by the tidal forces of the host galaxy \citep{darma+2021}.  Tidal captures are expected to occur more often in dwarf galaxies such as the LMC, rather than in the MW, due to a smaller relative velocity of star clusters \citep{vandenbergh+1996}. In spite of this, the encounter rate per cluster in the LMC is predicted to be quite low ($dN/dt\sim10^{-9}$ yr; \citealt{vallenari+1998}), making the possibility of an encounter highly unlikely for very young clusters.
%The encounter rate per cluster between systems with extremely high mass-ratio ($M_1/M_2\geq10^3$) is predicted to be quite low ($dN/dt\sim10^{-9}$yr; \citealt{vallenari+1998}). 
%Accordingly, very few cases of this kind are known observationally and no evidence of ongoing tidal captures has been collected so far, but only theoretical simulation studies are available \citep{delafuente+2014}. 
%These interactions should lead to the tidal destruction and/or the accretion of the lower-mass system (secondary) by the higher mass one (primary).  A stellar over-density in the region connecting the two clusters \citep{deoliveira+1998, leon+1999}, and a dynamical heating of the stars in the outer regions of the primary are also expected \citep{rao+1987, delafuente+2014}.

%% In the recent years, thanks to Hubble Space Telescope (HST), a superb
%% collection of high-resolution, deep images of star clusters in the LMC
%% and their surroundings have been secured, such a dataset can be
%% potentially exploited to discovery and characterize these elusive
%% systems.

In this paper we provide the first tentative evidence of an ongoing tidal capture of the small star cluster KMK88-10 by the massive globular cluster NGC 1835 in the LMC. This study is part of a project aimed at characterizing the dynamical age of old stellar clusters in the LMC \citep{ferraro+2019, lanzoni+2019}, by using the so-called ``dynamical clock'' (see \citealt{ferraro+2012, ferraro+2018, ferraro+2020, ferraro+2023, lanzoni+2016}). To this end, we secured a set of multiband Hubble Space Telescope (HST) images of NGC 1835, a very massive ($\sim6\times10^5 M_\odot$, \citealt{mackey+2003, mclaughlin+2005}), old ($t\sim13$ Gyr; \citealt{olsen+1998}) and metal-poor ([Fe/H]$\sim - 1.7$ dex; \citealt{mucciarelli+2021}) globular cluster lying close to the central bar of the LMC.
Near the border of the sampled field of view, a small and loosely populated star cluster was clearly visible: this is KMK88-10 (also named H88 120 and OGLE-CL LMC 74, see \citet{narloch+2022}).  It was first identified by \citet{pietrzynski+2000} and later studied by \citet{bica+2008, nayak+2016, narloch+2022}, who estimated a very low-mass ($\sim300 M_\odot$) and young age ($t\sim 500$ Myr). However, its characteristics have never been investigated in depth with adequate high-resolution and deep observations. Thanks to the acquired data-set, here we provide a detailed characterization of this stellar system and probe the possibility that it is experiencing a physical interaction with NGC 1835.  In agreement with the previous considerations about the encounter rate per cluster in the LMC, very few cases of this kind are known observationally, and no evidence of ongoing tidal captures between systems with very different ages and extremely high mass-ratio ($M_1/M_2\geq10^3$) has been collected so far. Only theoretical simulation studies are available \citep{delafuente+2014}, predicting that such interactions should lead to the tidal destruction and/or the accretion of the lower-mass system (secondary) by the higher mass one (primary).  A stellar over-density in the region connecting the two clusters \citep{deoliveira+1998, leon+1999}, and a dynamical heating of the stars in the outer regions of the primary are also expected \citep{rao+1987, delafuente+2014}.

The paper is organized as follows. In Section \ref{sec:reduction} we describe the data-set and the data reduction process. In Section \ref{sec:analysis} we present the determination of the age and structural parameters of the cluster. In Section \ref{sec:tidalcapture} we investigate the possibility of an ongoing tidal interaction between KMK88-10 and NGC 1835. The summary and discussion of the results are provided in Section \ref{sec:summary}.

\section{Data-set and data reduction}
\label{sec:reduction}
The adopted data-set consists of deep and high resolution images
obtained with the Wide Field Camera 3 (WFC3) onboard the HST under
program GO 16361 (PI: Ferraro). A total of 16 images were collected
with different filters:6 images in the ultraviolet (UV) filter F300X
with exposure times ranging from 917 s to 953 s, 6 in F606W with an
exposure time of 408 s, and 4 in F814W with exposure times ranging
from 630 s to 700 s.  For decontamination purpose, as part of the same
programme, a set of simultaneous parallel observations with the Wide
Field Camera (WFC) of the Advanced Camera for Surveys (ACS) have been
secured in the F606W and F814W filters to sample a region of the LMC
located at $\sim5\arcmin$ from the WFC3 pointing.   The data of
  NGC 1835 presented in this paper were obtained from the Mikulski
  Archive for Space Telescopes (MAST) at the Space Telescope Science
  Institute. The specific observations analyzed can be accessed via
  \dataset[DOI:
    10.17909/d4qh-wq06]{https://doi.org/10.17909/d4qh-wq06}.

The data reduction has been performed with the software DAOPHOT IV
\citep{stetson+1987}, following the recipes described in detail in
\citet{cadelano+2019, cadelano+2020a, cadelano+2020b}. The first step
consisted in the modeling of a spatially varying point spread function
(PSF) in each image and, for this purpose, approximately 200 bright,
isolated and well-distributed stars were selected and
analysed. Subsequently, we identified all the sources with a flux peak
above 5$\sigma$ from the level of the background, and we fitted the
PSF models previously found to all these detected sources. Using as
reference a master list including the stars identified in at least
half of the images acquired with the UV filter, we forced the fit of
the PSF model to the location of these sources in all the other
images, using DAOPHOT/ALLFRAME \citep{stetson+1994}. For each of the
identified stars, the magnitude values estimated in different images
were combined using DAOMATCH and DAOMASTER. The final catalogue
contains detector positions, instrumental magnitudes and photometric
errors for about 65000 sources. The magnitudes were then calibrated
onto the VEGAMAG photometric system, applying the appropriate aperture
corrections and zero points reported on the HST WFC3
website.\footnote{https://www.stsci.edu/hst/instrumentation/wfc3/data-analysis/photometric-calibration/uvis-photometric-calibration}.
The instrumental position were corrected for geometric distortion
effects using the correction coefficients quoted in
\citet{bellini+2011}. Finally, we transformed the corrected positions
to the absolute coordinate system ($\alpha, \delta$) by
cross-correlation with the Gaia DR3 catalogue
\citep{gaiacollaboration+2022}.

\section{ANALYSIS}
\label{sec:analysis}
\begin{figure}[ht!]
\centering
    \includegraphics[scale=0.3]{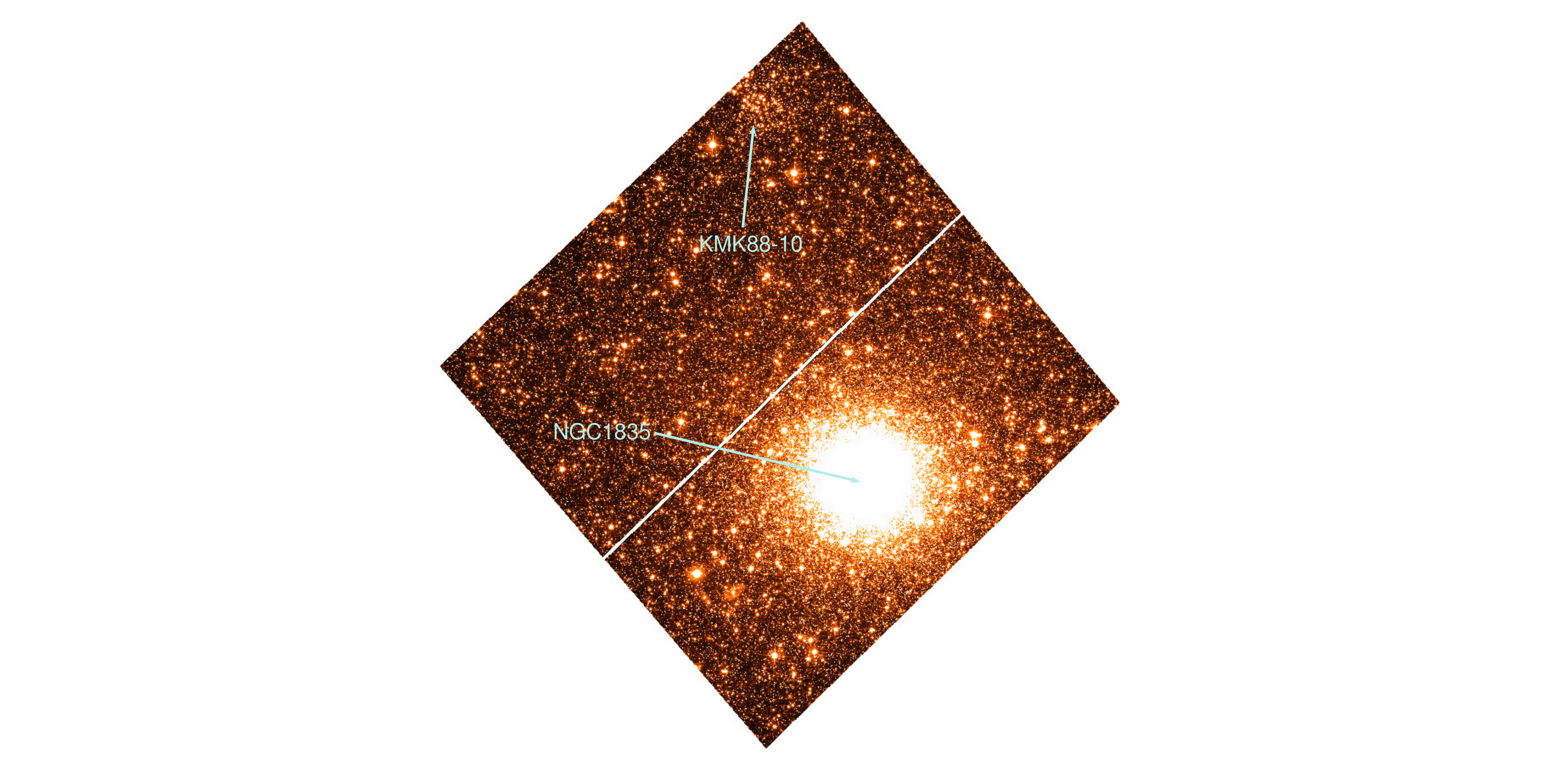}
    \caption{HST/WFC3 image in the F814W filter showing the positions
      of KMK88-10 and NGC 1835.  The field of view is $160\arcsec
      \times 160\arcsec$. North is up, and east is left.
\label{fig:image}}
\end{figure}

As anticipated in Sec.\ref{sec:intro}, the main target of the acquired
HST observations was NGC 1835. Hence one of the two chips of the WFC3
camera was roughly centered on the cluster center (see Figure
\ref{fig:image}). During the analysis we detected an over-density of
stars located at the northern edge of the field of view resembling a
small stellar cluster. By cross-correlating its position with the
SIMBAD catalog \citep{wenger+2000}, we identified this over-density as
the star cluster KMK88-10. While the photometric properties of the
main target (NGC 1835) will be presented in a companion paper (Giusti
et al., 2023, in preparation), here we discuss the characteristics of
KMK88-10 as derived from the analysis of this data-set. Figure
\ref{fig:image} shows one of the secured images in the F814W band,
with the positions of both KMK88-10 and NGC 1835
highlighted. Interestingly, the two stellar systems are separated by a
projected distance of about $130\arcsec$, which is only slighty larger
than the tidal radius of NGC 1835 of $\sim 105\arcsec$ (Giusti et al. 2023; \citet{mclaughlin+2005}).

\subsection{Decontamination of the color-magnitude diagram}
\label{subsec:decontamination}
\begin{figure}[ht!]
    \centering \includegraphics[scale=0.55]{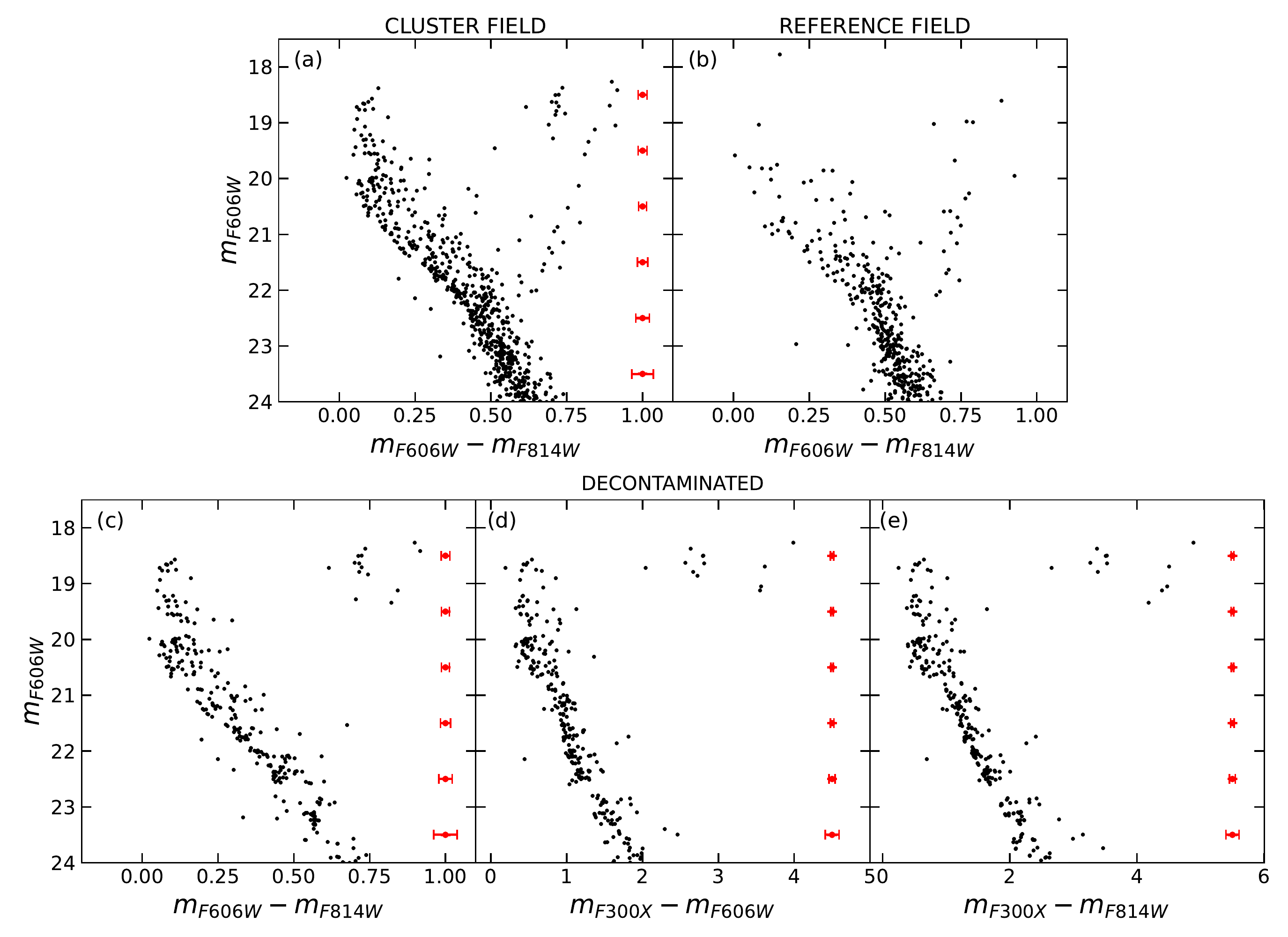}
    \caption{\emph{Panel (a)}: optical CMD of a circular area with a
      radius of $15\arcsec$ centered on KMK88-10. \emph{Panel (b):}
      optical CMD of a field region with an equal size area used for
      decontamination purposes. \emph{Panels: (c), (d), (e):}
      decontaminated CMDs of KMK88-10 in all the filter
      combinations. The typical photometric errors for different
      magnitude bins are shown in red.
\label{fig:deco}}
\end{figure}

The optical color-magnitude diagram (CMD) of a circular region with a radius of $15\arcsec$ centered on KMK88-10 is shown in Figure
\ref{fig:deco}(a). The radius was chosen as best compromise between
including most of the cluster stars and, at the same time, minimizing the number of field interlopers. Down to magnitude $m_{\rm F606W}=24$, this area includes 701 stars. The CMD shows a very extended main sequence (MS), ranging from $m_{\rm F606W}\sim18.4$ to $m_{\rm F606W}\sim25.5$. An extended and well-defined red giant branch (RGB) occupies the region at magnitude $18.2< m_{\rm F606W} < 22$ and color $0.5 < m_{\rm F606W}-m_{\rm F814W} < 0.9$. For comparison, in Figure \ref{fig:deco}(b) we show the CMD of a LMC field region of the same size, selected beyond the tidal radius of NGC 1835. From an immediate visual comparison between the cluster and the field CMDs, it appears evident that the extended RGB visible in panel (a) can be almost entirely attributed to the old field population, while the small clump of stars at $m_{\rm F606W}=18.5$ seems to be present only in KMK88-10. Also the brightest portion of the MS ($m_{\rm F606W} < 20$) appears to be as a component not present in the field.  The selected reference field area contains 409 stars down to magnitude $m_{\rm F606W}=24$. Thus, we can roughly estimate that the overall contribution of LMC field stars in the direction of KMK88-10 down to $m_{\rm F606W}=24$ is as large as $\sim 58\%$. This
clearly indicates that a proper decontamination of the CMD of KMK88-10 from the field contribution is mandatory to perform a detailed characterization of the cluster properties.

A decontamination through proper motions would be very challenging at the large distances of the LMC (see \citealp[e.g.][]{cadelano+2022a}), and in this case it is not feasible at all because of the lack of second-epoch of observations.  We therefore applied a statistical decontamination technique, following an approach similar to that proposed in \citet{milone+2018} (see also \citealp{dalessandro+2019}). For each star observed in the reference field, we removed one star from the cluster sample
%CMD. The identification of the potential interloper has been done
selected on the basis of its relative position in the CMD. More
specifically, for each $i^{th}$ star in the reference field, we
calculated its CMD distance ($d_i$)\footnote{The CMD distance between a star beloging to the cluster sample and the $i^{th}$ star of the reference field is defined as $d_i = \sqrt{(COL_c -COL_f)^2 + (MAG_c - MAG_f)^2}$, where $COL_c$ and $COL_f$ are the $(m_{\rm F606W}$-$m_{\rm F814W})$ colors of the stars in the cluster and field samples, respectively, while $MAG_c$ and $MAG_f$ are the corresponding values of their F606W magnitude.}  from all the stars belonging to the cluster sample. Then, we selected as most likely interloper stars in the cluster sample the objects having the lowest values of $d_i$. Thus, by construction, the procedure returns 409 field interlopers and 292 likely cluster member stars.  The entire process was repeated several times by using different reference fields of equal size.  All the decontaminated CMDs thus obtained are qualitatively in agreement, confirming that the adopted procedure provides reliable results. An example of decontaminated CMD obtained with this approach is shown in
Figure \ref{fig:deco}(c),(d),(e) for all the filter combinations.  As apparent, an extended MS survived the selection, indicating the existence of a young population with a MS turn-off point at $m_{\rm F606W}\approx19$.
Also a small clump of objects at the position of the red clump of this
young population remained in the cluster sample: see the stars located
at an average magnitude $m_{\rm F606W}\sim 18.5$ and color $(m_{\rm
  F606W}-m_{\rm F814W})\sim0.7$.
As expected, most of the stars along the RGB have been removed. The
only exception are four red objects with colors $ 0.8< (m_{\rm
  F606W}-m_{\rm F814W}) < 0.95$ and magnitudes $18 < m_{\rm F606W} <
19.5$, which most likely are residual field stars that survived the
decontamination process due to the low statistics along the bright
portion of the LMC RGB.

%\begin{figure}[ht!]
%    \centering
%    \includegraphics[width=18cm]{deco_12arcsec.pdf}
%    \caption{Decontaminazione con 12 arcsec
%    \label{fig:deco}}
%\end{figure}

\subsubsection{The cluster age}
\label{sec:age}
To estimate the cluster age, we compared the decontaminated CMD with a set of isochrones from the PARSEC database \citep{marigo+2017}. We
extracted isochrones at different ages (between 600 and 1000 Myr)
assuming a metallicity [Fe/H]$=-0.4$ dex, which is typical for the
young LMC stellar population \citep{mucciarelli+2008} and compatible
with that derived by \citet{narloch+2022}, who quote a value ranging
between $-0.52$ dex and $-0.44$ dex.  We also adopted a color excess of $E(B-V)=0.06$\footnote{We refer the reader to the forthcoming paper about NGC~1835 in which we will independently derive the $E(B-V)$ of the analyzed field of view.}.
 
\begin{figure}[ht!]
    \centering
    \includegraphics[scale=0.3]{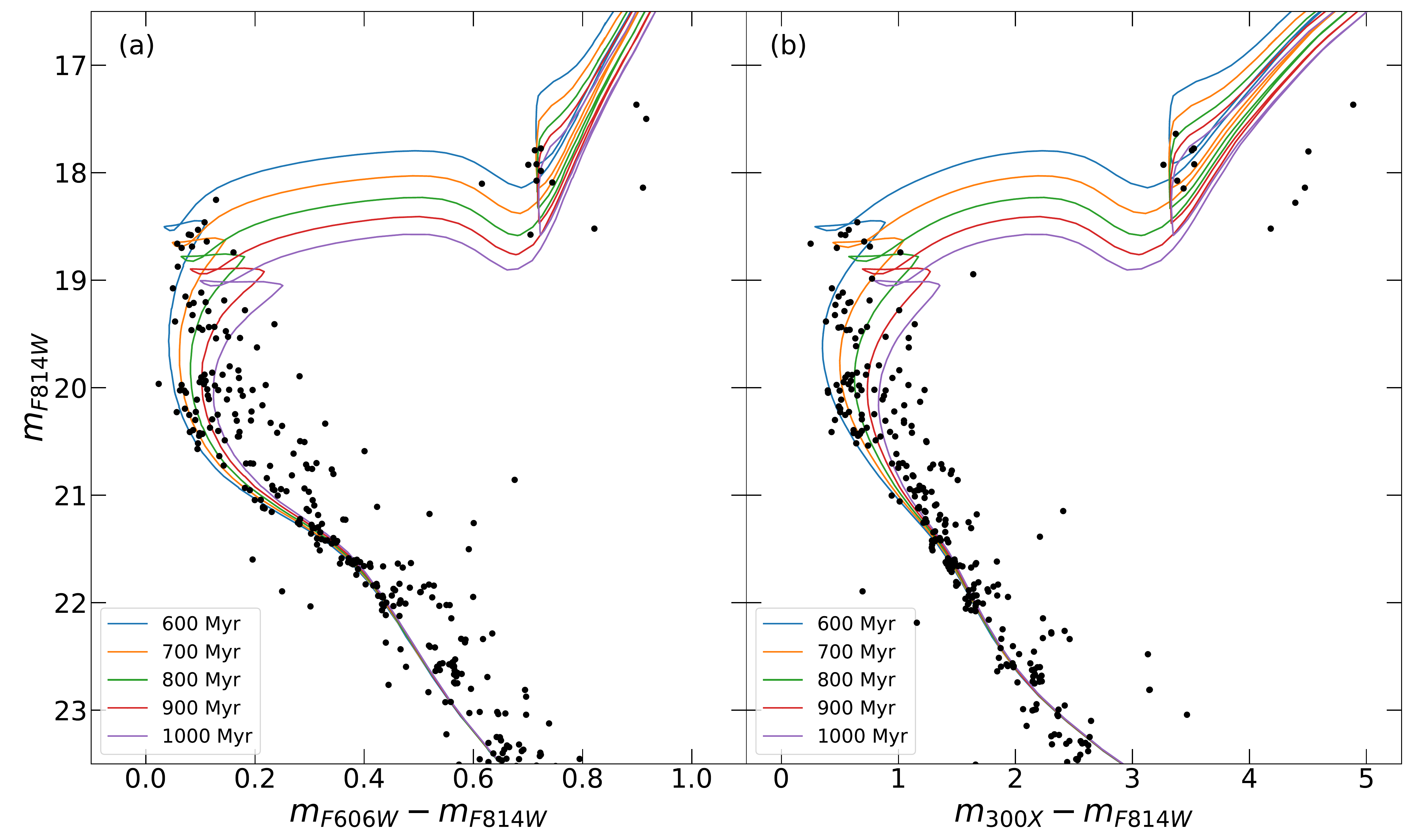}
    \caption{\emph{Panel (a)}: decontaminated optical CMD of KMK88-10
      (black dots) with a set of Parsec isochrones \citep{marigo+2017}
      computed for different ages superposed as colored lines (see the
      legend). The adopted metallicity is [Fe/H]$ = -0.4$ dex.  A
      distance modulus $(m-M)_0=18.48$ and a color excess $E(B-V)=
      0.06$ have been assumed to plot the isochrones in the observed
      CMD.  \emph{Panel(b)}: same as in Panel(a), but in the hybrid
      $m_{\rm F814W}$ vs. ($m_{\rm F300X}$-$m_{\rm F814W}$) CMD.
    \label{fig:iso}}
\end{figure}

Finally, we adopted  the LMC distance modulus $(m-M)_0=18.48$, corresponding to a distance of 49.6 kpc \citep{pietrzynski+2019}.  In Figure \ref{fig:iso} we show the decontaminated optical and hybrid ($m_{\rm F814W},m_{\rm F300X}$-$m_{\rm F814W}$) CMDs.  As can be seen, the MS turn-off region appears broadened in color, as commonly observed in young star clusters and likely due to the effects of stellar rotation \citep{mackey+2007, kamann+2020, martocchia+2023} and the presence of MS binaries. This hampers a solid determination of the cluster age, as rotating stars can be brighter or fainter than non-rotating
stars depending on their angular velocities and inclination angles
with respect to the observer.  However, the comparison with isochrones for  non-rotating stars, shows that the stellar population of this cluster is quite young, with an age between 600 Myr and 1 Gyr. In fact, isochrones in this age range nicely reproduce both the MS turn-off region, also accounting for its broadening in color, and the position of the red clump of stars.  These results are qualitatively in agreement with those independently obtained by \citet{narloch+2022}, who quote $t\sim 537 \pm 23 $ Myr.

\subsection{Surface brightness profile and structural parameters}
\label{subsec:sb profile}
\begin{figure}[ht!]
    \centering
    \includegraphics[scale=0.5]{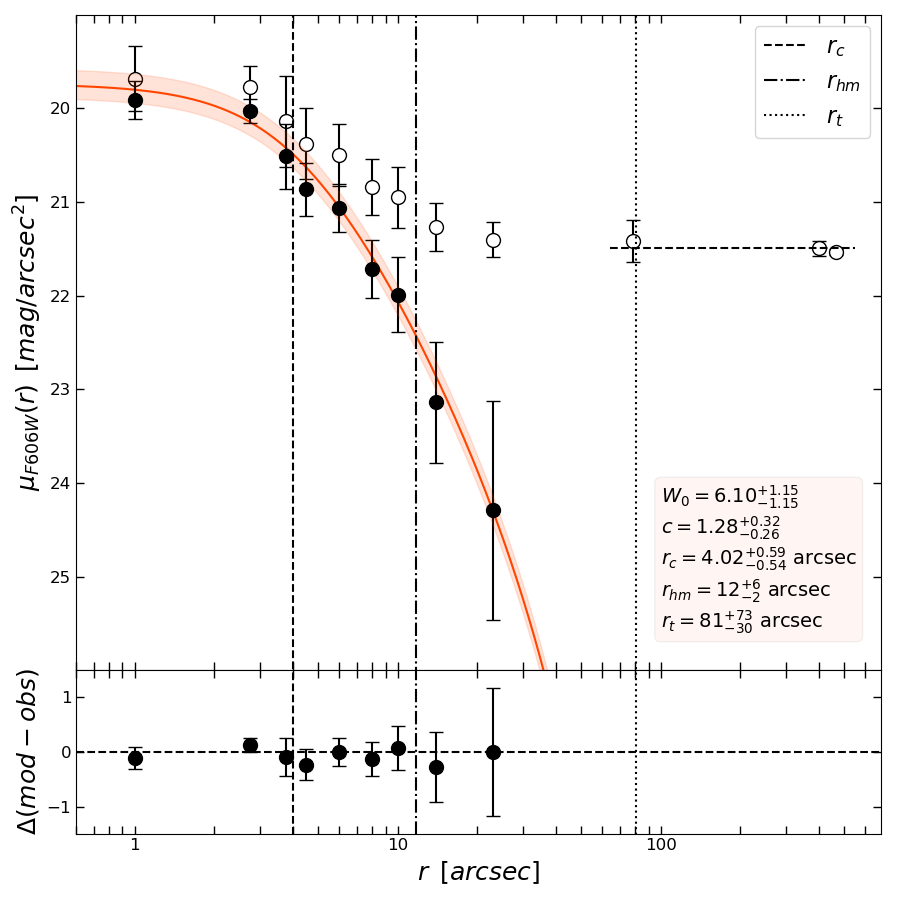}
    \caption{Surface brightness profile of KMK88-10 in the F606W
      filter. The open circles represent the observed measures, while
      the filled circles correspond to the cluster profile after
      subtraction of the LMC field contribution (horizontal dashed
      line). The red solid line shows the best-fit King model, with
      the corresponding structural parameter values labelled in the
      figure legend. The vertical lines mark the locations of the core
      radius (dashed line), half-mass radius (dot-dashed line), tidal
      radius (dotted line).
    \label{fig:sb}}
\end{figure}

The very low number of stars in this cluster hampers a
characterization of its density profile via number counts
\citep{lanzoni+2007, lanzoni+2010, lanzoni+2019, ibata+2009,
miocchi+2013, cadelano+2017}. Therefore, to constrain its structural
properties, we analyzed the surface brightness profile.  As a first
step, we divided the field of view into concentric radial rings
centered on the cluster gravitational center quoted in
\citet{nayak+2016,bica+2008}: $\alpha:05^h:05^m:05^s$, $\delta:
-69^{\circ}:24\arcmin:12\arcsec$.
%Each annulus was then divided into several sub-rings, depending on its 
%dimension.
Each ring was then divided in 4 sectors and the integrated surface
brightness of each sector was determined.  The average of the values
measured among the considered sectors was then adopted as integrated
surface brightness of each ring, while the standard deviation has been assumed as its uncertainty. In this procedure, the portions of the images included within the tidal radius of NGC 1835 were excluded. To properly estimate the background level, we extended the measurement of the surface brightness to the parallel images obtained with the ACS, in a region far distant from both the clusters. The resulting surface brightness profile obtained in the $m_{\rm F606W}$ filter is shown in Figure \ref{fig:sb} (open circles). The profile shows an almost constant value in the most central regions ($r<4\arcsec$), then it smoothly declines toward a plateau at distances $r > 20\arcsec$, where the contamination due to field stars becomes dominant. The average surface brightness of the last 3 points provided an estimate of the mean background surface brightness ($\mu_{\rm F606W} \sim 21.5$ mag
arcsec$^{-2}$) and this value was subtracted to all the observed
points to obtain the decontaminated profile of KMK88-10 (solid
circles).

The decontaminated profile has the typical shape of the King models
\citep{king+1966}, which are usually adopted to describe the surface
brightness of star clusters. Hence, to determine the structural
parameters of KMK88-10, we fitted the obtained profile with the family
of spherical, isotropic, single-mass King models. The comparison was
performed through a Monte Carlo Markov Chain fitting technique
following the prescriptions by \citet{raso+2020, cadelano+2022b,
deras+2023}. We assumed $\chi^2$ likelihood and flat priors for all
the fitting parameters. The resulting best-fit model is shown in
Figure \ref{fig:sb}. It turns out that the cluster is characterized by a King concentration index of about $c=1.3$, a central surface
brightness $\mu_{\rm F606W,0} = 19.8$ mag arcsec$^{-2}$, a core radius $r_c=4.0 \arcsec$, a half-mass radius $r_{hm}= 12\arcsec$, and a tidal radius $r_t= 81\arcsec$.
%Knowing the LMC distance the results can be expressed as $r_c= 0.94$ pc (compatible with the value obtained in \textcite{werchan+2011}), $r_{hm}=2.80$ pc and $r_t= 18.86$ pc. \\
By integrating the best-fit surface brightness profile, we obtained an integrated magnitude $m_{\rm F606W,int}=17.9$. This value, assuming the distance modulus and the color excess quoted above, corresponds to an absolute magnitude $M_{\rm F606W,int}=-0.75$ that, following the prescriptions by \citet{harris+2018}, corresponds to an absolute Johnson $V$-band magnitude $M_V=-0.71$. Finally, assuming a mass-to-light ratio $(M/L_V)\sim0.5-1$, as appropriate for such a young stellar population \citep{maraston+1998}, the total mass is $M\sim80M_\odot-160 M_\odot$, compatible with that of a low-mass open cluster \citep[e.g.][]{piskunov+2008}. This value is 2-4 times smaller than that derived by \citet{popescu+2012}, likely because of the different
data-sets, background decontamination procedure, and mass estimate technique adopted there.

\section{A tidal bridge between KMK88-10 and NGC 1835?}
\label{sec:tidalcapture}
%% Interestingly, the tidal radius of NGC 1835 ($r_t\sim105\arcsec$,
%% corresponding to $\sim 25$ pc) is only slighter smaller than its
%% projected distance from the gravitational center of KMK88-10
%% $d=130\arcsec$ ($\sim 31$ pc), for which we estimated a tidal radius
%% of $r_t\sim80\arcsec$ ($\sim 19$ pc).
 
Interestingly, the projected distance between the gravitational
centers of KMK88-10 and NGC 1385 ($d=130\arcsec$, corresponding to
$\sim 31$ pc) is smaller than the sum of the tidal radii of the two
clusters ($r_t\sim80\arcsec \sim 19$ pc for KMK88-10,
$r_t\sim105\arcsec \sim 25$ pc for NGC 1835).  Moreover, we find that
the two systems share similar bulk proper motions.  These have been
estimated from the Gaia Data Release 3 catalog
\citep{gaiacollaboration+2022} for the stars that survived the
decontamination procedure in KMK88-10, and for those located between
$15\arcsec$ and $40\arcsec$ from the center in NGC 1835 (this
selection is aimed to minimize the number of contaminating field
objects and, at the same time, to exclude low quality proper motion
measurements in the inner region of such a dense cluster).  We
obtained a final sample of 34 stars in KMK88-10 and 246 stars in NGC
1835, from which we calculated the average proper motion in right
ascension ($\mu_{\alpha}$) and in declination ($\mu_{\delta}$) of the
two clusters, finding:
%after applying a $3\sigma$-clipping rejection algorithm to both the
%samples.
$(\mu_{\alpha},\mu_{\delta})_{\rm KMK88-10}=(2.12 \pm 0.25, 0.02 \pm
0.27)$ mas yr$^{-1}$ for the former, and
$(\mu_{\alpha},\mu_{\delta})_{\rm NGC 1835}=(1.93 \pm 0.27, -0.08 \pm
0.30)$ mas yr$^{-1}$ for the latter. Although the statistic is poor,
the evidence of a comparable bulk proper motion, together with a small
projected distance on the sky, opens to the possibility that the two
clusters are gravitationally bound.

To further investigate this hypothesis,
we compared the observed separation with the Jacoby radius ($r_J$) and
the Roche-Lobe radius ($r_{\rm RL}$) of NGC 1835.  The former has been
calculated as \citep[][see also
  \citealt{dalessandro+2018}]{bertinvarri08}:
\begin{equation}
    r_{J} = \left(\frac{GM}{\xi\,\Omega^2}\right)^{1/3},
\end{equation} 
where $G$ is the gravitational constant, $M$ is the total mass of the
cluster, and $\Omega$ is its orbital frequency in the host
galaxy. Under the assumtion that the LMC potential is properly
described by a spherical Plummer model with characteristic radius $b =
2.6$ kpc \citep[see, e.g.,][]{bekki+2005}, the dimensionless parameter
$\xi$ can be calculated as:
\begin{equation}
    \xi(R_0) = \frac{3R_0^2}{b^2+R_0^2},
\end{equation}
where $R_0$ is the radius of the circular orbit (i.e., the
galactocentric distance of the cluster).  Assuming a total mass
$M=6\times 10^5 M_\odot$ \citep{mackey+2003, mclaughlin+2005}, a
galactocentric radius $R_0= 1.56$ kpc \citep{mclaughlin+2005} and an
orbital frequency $\Omega= 0.03$ km s$^{-1}$ pc$^{-1}$
\citep{dalessandro+2018}, we find that the Jacobi radius of NGC 1835
is $r_J\sim 155$ pc, which suggests that KMK88-10 is totally contained
(at least in projection) within its gravitational sphere of influence.
Finally, the Roche-Lobe radius of NGC 1835, calculated following
\citet{eggleton+1983} and assuming a mass ratio of $\sim1000$, turns
out to be $\sim0.8$ times the separation between the two systems
($r_{\rm RL}\sim 25$ pc), thus indicating that KMK88-10 extends within
it.

\begin{figure}[ht!]
    \centering \includegraphics[scale=0.65]{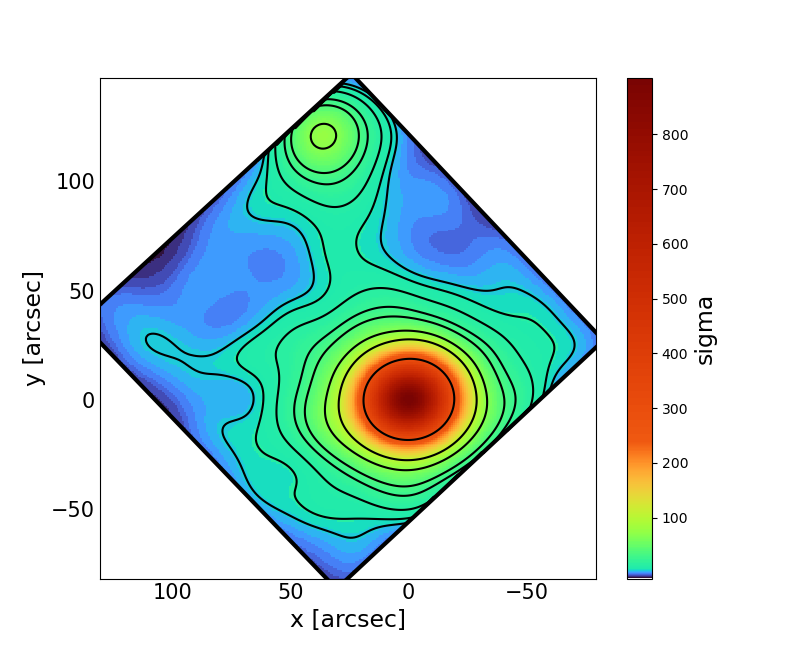}
    \caption{Surface density map of the field sampled by the
        discussed HST/WFC3 observations, with the same size and
        orientation of Fig. \ref{fig:image}.  Different colors
        correspond to different values of significance ($\sigma$)
        above the background level (see the side color-bar).  The
        isodensity contour levels (black lines) range from $\sim
        4\sigma$ to $\sim 300\sigma$, with irregular steps.  The
        highest density peak corresponds to the location of NGC 1835,
        while the northen overdensity is KMK88-10.  The presence of a
        bridge between the two systems is also apparent.}
\label{fig:2dmap}
\end{figure}

Although both $r_J$ and $r_{\rm RL}$ have been compared with a
projected (rather than 3D) distance, it is definitively worth
investigating the intriguing possibility that the proximity of these
two systems is not merely due to a chance coincidence, but to a
physical connection. The lack of tidal tails in the observed surface
brightness profile of KMK88-10 is not conclusive in this sense,
because the largely dominant field contribution in the external region
of the system could easily hide this feature. Thus, to test this
hypothesis we created a 2D density map where the severe contamination
from field objects is minimized. To this end, we first drew selection
boxes enclosing the MS and post-MS stars of KMK88-10 in its
decontaminated CMD (bottom panel of Figure \ref{fig:deco}), and three
additional selection boxes corresponding to the MS, RGB and horizontal
branch sequences of NGC 1835. Then, we selected the stars located
across the whole field of view and falling within one of these boxes
in the observed CMD.  To drastically reduce the contamination from
bright field stars, the sample was further refined including only
stars with $m_{\rm F606W}<20.2$.  The 2D density map of this sample
was created by gridding the field of view in $1\arcsec \times
1\arcsec$ cells, and applying a Gaussian kernel to obtain a smoothed
distribution.  The result is shown in Figure \ref{fig:2dmap},
with different colors and lines corresponding to different
  values of significance ($\sigma$) above the background level. The
  latter is computed as the mean value of the smoothed density
  distribution in a region far from both clusters, toward the east
  corner of the sampled field of view (see Fig. \ref{fig:image}),
  while $\sigma$ corresponds to its standard deviation. The map
clearly shows the presence of two overdensities in north and
south-west directions, centered on KMK88-10 and NGC 1835,
respectively. Interestingly, the KMK88-10 density distribution appears
to be non spherical and elongated toward the direction of NGC
1835. Indeed, the isodensity countours have significant ellipticity,
with the major axis directed toward the center of NGC
1835. Furthermore, a small overdensity of stars is present in the
region connecting the two clusters, creating a sort of bridge between
the two systems. Despite the small number of stars and some residual
contamination from field objects, this feature suggests a possible
physical connection between the two clusters.

\begin{figure}[ht!]
    \centering \includegraphics[scale=0.5]{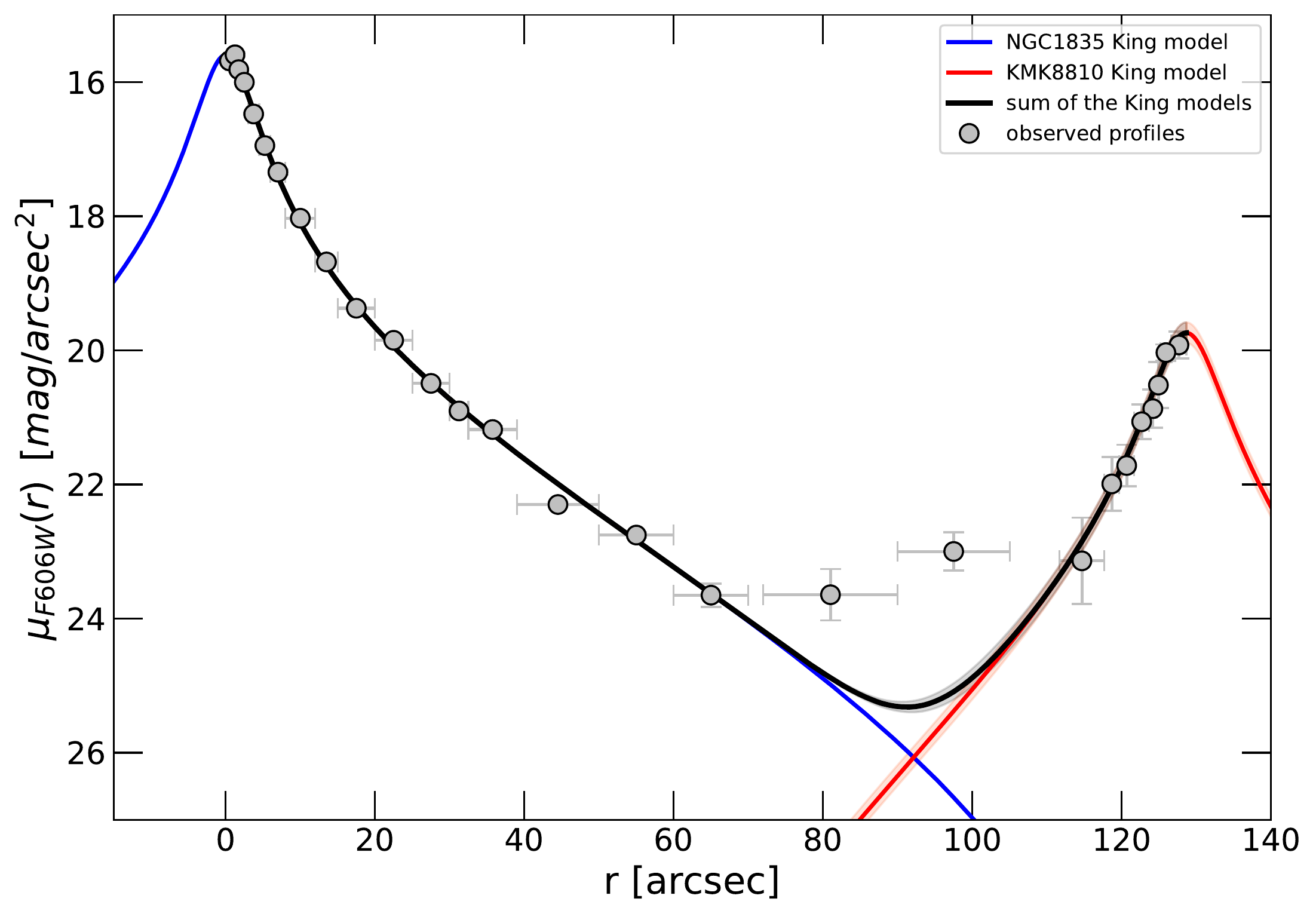}
    \caption{Decontaminated surface brightness profile (grey circles)
      of NGC 1835 (Giusti et al., in prep.), KMK88-10 (the same as in
      Fig. \ref{fig:sb}), and the bridge region between them, as a
      function of the distance from the center of the globular
      cluster. The blue and red lines are the best-fit King models to
      the profiles of NGC 1835 and KMK88-10, respectively.  The black
      line is the sum of the two King models.}
\label{fig:sommaprofili}
\end{figure}

In principle, however, the observed intracluster overdensity could
also be due to the mere superposition of the two individual density
profiles.  To investigate this possibility we determined the surface
brightness profile of NGC 1835 using the same method adopted for
KMK88-10 (see Section \ref{subsec:sb profile}), and we measured the
surface brightness distribution in two rectangular regions covering
the ``brigde area'' between the two systems. Since the field level
found for NGC 1835 is equal to that observed for KMK88-10 ($\mu_{\rm
  F606W}\sim 21.4$ mag/arcsec$^{-2}$; see the horizontal dashed line
in Fig. \ref{fig:sb}), this has been used to decontaminate all the
observed surface brightness values from the LMC contribution. The
results, as a function of the distance from the center of NGC 1835,
are shown as grey circles in Fig.\ref{fig:sommaprofili}.  The red line
is the best-fit King model to the surface brightness profile of
KMK88-10 (the same as in Fig. \ref{fig:sb}), while the blue line is
the same but for NGC 1835 (Giusti et al., in prep.). The black line is
the sum of the two profiles.  As apparent, only a very small
overdensity of $\mu_{\rm F606W}\sim25$ is expected at $r\sim90\arcsec$
from the superposition of the surface brightness profiles of the two
clusters, while the value measured along the bridge at the same
distance is $\sim 2$ magnitudes brighter. This indicates that the
observed intracluster bridge could be indeed the result of an ongoing
interaction between the two systems, where the small cluster KMK88-10
is being tidally captured by the massive globular cluster NGC 1835.
  %% As a final note, we stress that, although for KMK88-10 we adopted
  %% the average LMC distance modulus of 18.48, the results in terms of
  %% cluster age and intracluster density would not change if we assumed
  %% $(m-M)_0= 18.62$, which is the value estimated for NGC 1835 (Giusti
  %% et al., in prep.). FORSE QUESTO LO POSSIAMO OMETTERE QUA E DIRLO
  %% LA`..
Unfortunately, KMK88-10 is adiacent to the northen edge of the
  sampled field of view, thus preventing any detection of the trailing
  tail expected in the case of a tidal distrotion effect. On the other
  hand, we verified that the bridge is not visible in the density map
  obtained from stars belonging to NGC 1835 alone (i.e., selected
  along the MS and RGB of the globular cluster, at magnitudes $m_{\rm
    F606W}>20$, where the contribution from KMK88-10 stars is totally
  negligible). This indicates that the bridge is primarily composed of
  KMK88-10 stars that are gravitationally pulled in by NGC 1835.

\section{Summary and conclusions}
\label{sec:summary}
We used high-resolution HST/WFC3 images to analyze the stellar
population of KMK88-10, a small star cluster located in the LMC. This system lies at an angular separation of approximately $130\arcsec$ in sky from the massive and old globular cluster NGC 1835. The adopted data-set allowed us to construct the deepest CMD obtained so far for KMK88-10, properly decontaminated from the strong contribution of LMC field stars. In turn, this allowed a detailed characterization of the stellar population and structural properties of the system.

The decontaminated CMD clearly shows the presence of a young (600-1000 Myr old) stellar population. We detected a significant broadening of the MS turn-off region, suggesting that this stellar system is affected by the same phenomenon observed in many other young LMC clusters, possibly connected to stellar rotation.  We also analyzed the surface brightness profile of the cluster and we fit it with a King model. Under the assumption of a distance modulus $(m-M)_0=18.48$, we found
%% a core radius of $4\arcsec$, a half-mass radius of $12\arcsec$ and a
%% concentration parameter $c\sim1.3$, corresponding to a truncation
%% radius of $\sim81\arcsec$. The integration of the best-fit brightness
a core radius of 0.97 pc, a half-mass radius of 2.89 pc and a
concentration parameter $c\sim1.3$, corresponding to a truncation
radius of about 19 pc.  The integration of the best-fit surface
brightness profile provides an integrated $V$-band absolute magnitude of $M_V=-0.71$, corresponding to a mass $M\sim80-160 M_\odot$ for a mass-to-light ratio $(M/L_V)\sim0.5-1$.  The main physical parameters of the cluster are summarized in Table \ref{table:1}.

At the LMC distance, the small angular separation between KMK88-10 and
NGC 1835 ($130\arcsec$) corresponds to a projected physical distance
of only $\sim 30$ pc,   which is smaller than the sum of the two
  tidal radii, and than the Jacobi radius of the globular cluster. In
  addition, KMK88-10 turns out to extend within the Roche-Lobe radius
  of NGC 1835.  Although the three dimensional distance is likely
larger than the projected one, the observed proximity promoted the
investigation of a possible physical link between the two systems.
With this purpose, we constructed the 2D surface density map of the
likely cluster members surveyed in the WFC3 field of view, finding
that KMK88-10 has an elongated structure toward the direction of NGC
1835 (see Fig. \ref{fig:2dmap}).  Furthermore, an over-density
  between the two clusters significantly exceeding the superposition
  of the two surface brightness profiles has been detected,
suggesting the presence of a tidal bridge between the two systems.  If
confirmed, this would be the first evidence of a globular cluster
caught in the act of tidally capturing a young and low-mass star
cluster.  According to our mass estimate for KMK88-10 and the value
quoted by \citet{mackey+2003, mclaughlin+2005} for NGC 1835, the two
interacting stellar systems would have an extremely high mass ratio
($M_{\rm NGC 1835}/M_{\rm KMK88-10}\sim 6\times 10^3$).  As discussed
above, such events are supposed to be quite rare and have significant
impact on the evolution and the observed characteristics of massive
star clusters. Indeed, cluster mergers could explain the
  properties of massive and dynamically complex systems
  \citep[e.g.][]{baumgardt+2003, bruns+2011} and could provide an
  explanation to the presence of the light-element complexities in LMC
  intermediate-age clusters \citep{hong+2017} and multiple iron
  distribution in some old clusters \citep{gavagnin+2016}.
%In fact, cluster mergers could explain the properties of massive and dynamically complex systems \citep[e.g.   []{baumgardt+2003, bruns+2011} and could provide an explanation to the presence of some of the observed chemical complexities \citep[e.g.][]{gavagnin+2016, hong+2017}. 
The final fate of this interaction should lead to the tidal destruction of KMK88-10 in a time scale that depends on its orbit, possibly leaving an observable kinematical signature due to the dynamical heating of the stars in the outer regions of NGC 1835.  This could be observable through a rise in the stellar velocity dispersion in the cluster outskirts. Hence, spectroscopic observations aimed at measuring the stellar velocity dispersion profile of NGC 1835 might shed light on this, and help confirming the tidal capture scenario.

%Although such a phenomenon is expected to be extremely rare, dwarf galaxies, such as the Magellanic Clouds, because of their small velocity dispersion, represent the ideal environment where to search and characterize this type of cluster interactions. This work demonstrates that the large amount of HST archival data of star clusters in the Magellanic Clouds provide an ideal opportunity to discover and characterize low mass clusters which might be elusive in ground based observations. 

\begin{table}\label{table}
\centering
\begin{tabular}{l l}
\hline
\hline
Parameter & Estimated value \\
\hline
Center of gravity & $\alpha_{J2000}$ = $5^h05^m14^s$\\
 & $\delta_{J2000}$ = $-69^{\circ}22' 12''$ \\
Age &  $t=600$ Myr - 1 Gyr \\
King concentration & $c=1.28^{+0.32}_{-0.26}$\\
Core radius & $r_c = 4.02^{+0.59}_{-0.54}$ arcsec (0.97 pc)\\
Half mass radius & $r_{hm} = 12^{+6}_{-2}$ arcsec (2.89 pc)\\
Tidal radius & $r_t = 81^{+73}_{-30}$ arcsec (19.49 pc)\\
Integrated $V$ magnitude  & $M_V = -0.71$ \\
Mass & $M = 80- 160  M_\odot$\\
\hline
\end{tabular}
\caption{Summary of the main properties of KMK88-10.}
\label{table:1}
\end{table}

%% \begin{table}\label{table}
%% \centering
%% \begin{tabular}{l l}
%% \hline
%% \hline
%% Parameter & Estimated value \\
%% \hline
%% Center of gravity & $\alpha_{J2000}$ = $5^h05^m14^s$\\
%%  & $\delta_{J2000}$ = $-69^{\circ}22' 12''$ \\
%% Age &  600 Myr - 1 Gyr \\
%% Core radius ($r_c$) & $4.02^{+0.59}_{-0.54}$ arcsec (0.97 pc)\\
%% Half mass radius ($r_{hm}$) & $12^{+6}_{-2}$ arcsec (2.89 pc)\\
%% Tidal radius ($r_{t}$)& $81^{+73}_{-30}$ arcsec (19.49 pc)\\
%% Integrated V magnitude ($M_{V,int}$) & $-0.71$ \\
%% Mass & $80 \, M_{\odot} - 160 \, M_{\odot}$\\
%% \hline
%% \end{tabular}
%% \caption{Summary of the main properties of KMK88-10.}
%% \label{table:1}
%% \end{table}

\begin{acknowledgments}
This work is part of the project {\it Cosmic-Lab} ({\it "Globular Clusters as Cosmic Laboratories"}) 
 at the Physics and
Astronomy Department ``A. Righi'' of the Bologna University (http://
www.cosmic-lab.eu/ Cosmic-Lab/Home.html). The research was funded by
the MIUR through the PRIN-2017 grant awarded to the project
Light-on-Dark (PI:Ferraro) with contract PRIN-2017K7REXT.
\end{acknowledgments}

\vspace{5mm}
\facilities{HST(WFC3), HST(ACS)}

\bibliography{bibliografia}
\bibliographystyle{aasjournal}

\end{document}